\documentclass[pra,aps,amsfonts,amsmath,superscriptaddress,%
  twocolumn,showpacs,floatfix]{revtex4}
\usepackage{amsfonts}
\usepackage{amsmath}
\usepackage{bm}
\usepackage{epsfig}
\usepackage{mathrsfs}

\begin{document}

 \title {\bf Bubbles in $^{34}$Si and $^{22}$O?}
\author{M. Grasso}
\address{Institut de Physique Nucl\'eaire,
 Universit\'e Paris-Sud, IN2P3-CNRS, F-91406 Orsay Cedex, France}
\author{L. Gaudefroy}
\address{CEA/DAM Ile de France, Service de Physique Nucl\'eaire,
Bruy\`ere-le-Ch\^atel, F-91297 Arpajon Cedex, France}
\author{E. Khan}
\address{Institut de Physique Nucl\'eaire,
 Universit\'e Paris-Sud, IN2P3-CNRS, F-91406 Orsay Cedex, France}
\author{T. Nik\v si\' c}
\address{Physics Department, Faculty of Science, Zagreb University, 10000
Zagreb, Croatia}
\author{J. Piekarewicz}
\address{Florida State University, Tallahassee, Florida 32306, USA}
\author{O. Sorlin}
\address{Grand Acc\'elerateur National d'Ions Lourds (GANIL),
CEA/DSM-CNRS-IN2P3, Bd Henri Becquerel, BP 55027, F-14076 Caen Cedex 5,
France}
\author{N. Van Giai}
\address{Institut de Physique Nucl\'eaire,
 Universit\'e Paris-Sud, IN2P3-CNRS, F-91406 Orsay Cedex, France}
\author{D. Vretenar}
\address{Physics Department, Faculty of Science, Zagreb University, 10000
Zagreb, Croatia}

\begin{abstract}
 Bubble nuclei are characterized by a depletion of their central
 density.  Their existence is examined within three different
 theoretical frameworks: the shell model as well as non-relativistic
 and relativistic microscopic mean-field approaches. We propose
 $^{34}$Si and $^{22}$O as possible candidates for proton and neutron
 bubble nuclei, respectively. In the case of $^{22}$O, we observe a
 significant model dependence, thereby calling into question the
 bubble structure of $^{22}$O. In contrast, an overall agreement
 among the models is obtained for $^{34}$Si. Indeed, all models
 predict a central proton density depletion of about 40\%. This
 result provides strong evidence in favor of a proton bubble in
 $^{34}$Si.
\end{abstract}

\vskip 0.5cm
\pacs{21.10.Ft, 21.10Re, 21.60.Jz, 25.30.Bf,25.70.De}
\maketitle


\section{Introduction}

The {\sl ``bubble''} structure of atomic nuclei is characterized by a
depleted central density. Although it is somewhat unexpected that a
{\sl ``hole''} can be made in a nuclear system where nuclear forces
generate a saturation density ($\rho_0 \sim$ 0.16 fm$^{-3}$), this
phenomenon has been discussed for many decades. Indeed, the
possibility of bubble nuclei started with the pioneering work of
Wilson in the 40s~\cite{wilson} who studied the low-energy excitations
of a thin spherical shell, up to the first microscopic calculations of
Campi and Sprung in the 70s~\cite{campi}. More recently, bubbles have
been discussed in superheavy and hyperheavy
nuclei~\cite{bender,decharge}. The promise of producing more exotic
nuclei with the new generation of RIB facilities has revived the
interest in this subject.

Due to the absence of a centrifugal barrier, $s$-orbitals have radial
distributions peaked in the interior of the nucleus, with their
corresponding wave function extending further into the surface
depending on the number of nodes. In contrast, orbitals with non-zero
angular momenta are suppressed in the nuclear interior and do not
contribute to the central density. Therefore, any vacancy of
$s$-orbitals is expected to produce a depletion of the central
density. By using electron scattering from $^{206}$Pb and $^{205}$Tl
up to large momentum transfers, the radial distribution of the 3$s$
proton orbital was experimentally mapped and shown to closely resemble
the one predicted by an independent particle model.  The agreement
extends from the center of the $^{206}$Pb nucleus all the way to the
surface and reproduces accurately the nodal structure of the wave
function~\cite{cave,pandha}. Differences in the charge density between
$^{206}$Pb and $^{205}$Tl revealed that about 80\% of the proton
removal strength came from the $3s$ state, thereby leading to a
significant depletion of the proton density in the nuclear interior.
Specifically, the depletion fraction, defined as
\begin{equation}
 \label{eq:fraction}
  F \equiv \frac{\rho_{\rm max}\!-\!\rho_{\rm c}}{\rho_{\rm max}}\;,
\end{equation}
amounts to $F\!=\!11(2)$\%. Note that here $\rho_{\rm c}$ and
$\rho_{\rm max}$ represent the values of the central and
maximum charge density in $^{205}$Tl, respectively. Yet the
small energy difference between the $3s_{1/2}$ and the $2d_{3/2}$
proton orbitals plus the coupling of the $3s_{1/2}$ proton to
collective excitations in $^{206}$Pb, yield a proton hole strength in
$^{205}$Tl that is shared among the $3s_{1/2}$ and $2d_{3/2}$
orbitals, with the former carrying about 70\% of the strength and the
latter the remaining 30\%.  Consequently, the central depletion in
$^{205}$Tl relative to $^{206}$Pb is not as large as if the full hole
strength would have been carried by the $3s$ orbital. Using similar
arguments, the depletion at the center of $^{204}$Hg is not expected
to be very large, as the two-proton hole strength will be again shared
among the $3s_{1/2}$ and $2d_{3/2}$ orbitals. Therefore, the search
for the best bubble candidates should be oriented towards nuclei with
an $s$ orbital well separated from its nearby single-particle states
and where correlations are weak. This latter feature arises mainly for
nuclei located at major shell closures.

Recently, the formation of a proton bubble due to the depletion of
the $2s_{1/2}$ orbital was investigated in $^{46}$Ar~\cite{tod04,kha08}
and in the very neutron-rich Ar isotopes~\cite{kha08}. In $^{46}$Ar
the proton $2s_{1/2}$ and $1d_{3/2}$ orbitals are almost degenerate.
Thus, as in the case of $^{206}$Pb, pairing correlations will lead
to a significant occupancy of the $2s_{1/2}$ orbital~\cite{Gaud07},
thus weakening the bubble effect. This weakening will continue to
hold for any $N=28$ isotone between $Z=20$ and $Z=14$ as long as
the $2s_{1/2}$ and $1d_{3/2}$ orbitals remain degenerate, as shown
for instance in Fig.~3 of Ref.~\cite{gade}. For very neutron-rich
Ar isotopes, such as $^{68}$Ar, the $s_{1/2}$ proton orbital is
predicted to move significantly above the $d_{3/2}$ state, hindering
the role of pairing correlations ~\cite{kha08,gra07}. Unfortunately,
the production of this exotic nucleus is far beyond the present and
near-future capabilities of RIB facilities.

A more suitable region of the chart of the nuclides to
search for a proton bubble is that of the $N=20$ isotones.
Between $Z=20$ and $Z=16$ the $s_{1/2}$ orbital is located
about 6.5~MeV above the $d_{5/2}$ orbital and about 2.5~MeV
below the $d_{3/2}$ orbital, thereby forming two subshell
closures at $Z=14$ and $Z=16$, respectively~\cite{Prog08}.
In addition, the $N=20$ shell closure is rigid enough to
hinder significant coupling to collective states. Assuming
a sequential filling of proton orbitals, the $2s_{1/2}$
orbital should be completely empty in $^{34}$Si while fully
filled in $^{36}$S.  This may lead to an important change
in the proton density distribution between $^{36}$S and
$^{34}$Si making $^{34}$Si an excellent candidate for a
bubble nucleus.  Concomitantly, both Skyrme and Gogny
Hartree-Fock-Bogoliubov models predict a spherical shape
for $^{34}$Si \cite{yoshida,b3}. Other possible candidates
in the Si-isotopic chain, such as $^{28}$Si and $^{42}$Si,
are not optimal as they are deformed~\cite{28Si,Bast07}.
For these nuclei several correlations hinder the
development of a bubble.  The mirror system of the
({$^{36}$S, $^{34}$Si) system, ($^{36}$Ca,
$^{34}$Ca), could not be studied at present because the
$^{34}$Ca nucleus has so far not been observed.

A neutron bubble may be found in the oxygen chain, where
large $N=14$ (between $d_{5/2}$ and $s_{1/2}$) and $N=16$
(between $s_{1/2}$ and $d_{3/2}$) subshell gaps of about
4.2~MeV\cite{Sta04,Sch07} and 4~MeV~\cite{Ele07},
respectively, have been determined. Combined to the
large proton gap at $Z=8$, the $^{22}$O~\cite{Thi00,
Sta04,Bec06} and $^{24}$O~\cite{Sch07,Ele07,Hoff08}
theferore behave as doubly magic nuclei. In this case the
change in the occupancy of the $2s_{1/2}$ neutron orbital
will occur between $^{22}$O and $^{24}$O, making $^{22}$O a
good candidate for a neutron bubble nucleus.

The present article aims at determining whether $^{34}$Si
and $^{22}$O could be considered \textbf{as} good proton
and neutron bubble nuclei, respectively. Various
theoretical approaches will be employed to test the
robustness of the results. In Sec.~II these nuclei are
analyzed in terms of shell-model calculations so that
occupancies of the proton and neutron orbitals may be
determined. In Sec.~III we start by addressing (in III-A)
the role of pairing correlations in mean field approaches,
followed then with results on nucleon density profiles
obtained from: (i) non-relativistic Hartree-Fock (HF) and
Hartree-Fock-Bogoliubov (HFB) (in III-B) and (ii)
relativistic mean field (RMF) and relativistic
Hartree-Bogoliubov (RHB) (in III-C) microscopic
calculations. Comparison to experimental data will be made
whenever possible. Conclusions are drawn in Sec. IV.

\section{Shell-Model predictions}
The occurrence of bubbles in nuclei, as previously defined,
is directly linked to the occupancy of $s_{1/2}$ orbitals.
For both bubble candidates under study in this article,
$^{22}$O and $^{34}$Si, experimental values for the
occupancies are not yet available. Thus, we rely hereafter
on Shell Model (SM) calculations that are known to give
reliable estimates in neutron-rich nuclei occupying
the $sdpf$ shells~\cite{NOWACKI}.

Shell-model calculations have been performed with the \textsc{antoine}
code~\cite{ANTOINE1,ANTOINE2} using the SDPF-NR
interaction~\cite{NOWACKI}. The full $sd$ valence space was considered
for both proton and neutron excitations for each nucleus under
consideration. Moreover, in the calculations for $^{36}$S and
$^{34}$Si, $4p4h$ neutron excitations were allowed from the $\nu
d_{3/2}$ to the $\nu f_{7/2}$ orbitals.

\subsection{Neutron Bubble: $^{24}$O and $^{22}$O}
\begin{table}[!t]
  \caption{\label{OCCUP_SM_O} Ground state occupation numbers of
    neutron orbitals obtained in the SM framework for $^{24}$O
    and $^{22}$O.}
    \begin{ruledtabular}
      \begin{tabular}{cccc}
        Orbital         & $^{24}$O  & $^{22}$O  \\
       \hline
        $\nu 1 d_{5/2}$ &   5.75    &   5.38    \\
        $\nu 2 s_{1/2}$ &   1.89    &   0.34    \\
        $\nu 1 d_{3/2}$ &   0.36    &   0.28    \\
      \end{tabular}
    \end{ruledtabular}
\end{table}

The mean occupation numbers of neutron orbitals deduced
from SM calculations are reported in
Table~\ref{OCCUP_SM_O}. Contrary to naive expectations, the
mean number of neutrons removed from the $\nu s_{1/2}$
orbital while moving from $^{24}$O to $^{22}$O amounts to
only 1.55 (or 78\% in fraction of shell
coccupancy). The remaining neutron strength is removed
from the $\nu d_{5/2}$ orbital (0.37 or 18\%) and the $\nu
d_{3/2}$ orbital (0.08 or 4\%). These numbers suggest that
the depletion of the neutron density in the nuclear
interior of $^{22}$O relative to $^{24}$O may not be as
large as required for the formation of a bubble structure.
To quantify the size of the neutron hole the radial
dependence of the neutron wave functions in $^{22,24}$O
have been calculated using a standard Woods-Saxon
potential~\cite{BOHR}, the parameters of which are as
follows: $V_0$~=~-50~MeV, $a$~=~0.65~fm and
$r_0$~=~1.25~fm. The neutron densities displayed in
Fig.~\ref{FIGURE1} are the sum of the squared wave
functions calculated with the Woods-Saxon potential
(without spin-orbit term) weighted by the occupation
numbers obtained within the SM framework. The orbitals
occupied by the core neutrons (namely, the $1\nu s_{1/2}$,
$1\nu p_{3/2}$ and $1\nu p_{1/2}$ orbitals) are assumed to
be completely filled.
\begin{figure}[!b]
\includegraphics[width=8.6cm]{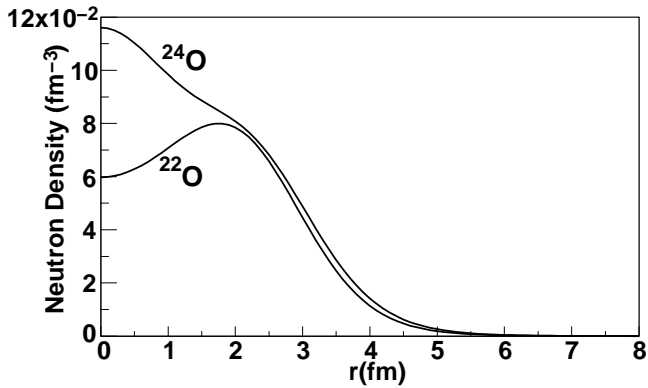}
\caption{\label{FIGURE1} Neutron density of $^{24}$O and $^{22}$O
  obtained using the occupation numbers of neutron orbitals resulting
  from SM calculations and radial wave functions obtained with a Woods-Saxon
  potential (see text).}
\end{figure}
The figure shows the variation of the neutron density
between $^{24}$O and $^{22}$O. In particular, the predicted
depletion fraction, as defined in Eq.~(\ref{eq:fraction}),
is found to be F~=~25\%. As will be shown in the following
section, the depletion in $^{22}$O is considerably smaller
than in $^{34}$Si. One notices that in the lighter Oxygen
isotopes this depletion factor is not expected to
increase as the neutron $1\nu d_{5/2}$ orbital is
depleted in concert, thus lowering the maximum value of
the density around the surface of the nucleus. It
follows that the relative difference of the density in the
vicinity of the surface and at the interior of the nucleus
is also reduced.

\subsection{Proton Bubble: $^{36}$S and $^{34}$Si}

The mean occupation numbers of the proton $1 d_{3/2}$, $2 s_{1/2}$ and
$1 d_{5/2}$ orbitals in $^{36}$S have been obtained from the
$^{36}$S(d,$^3He$)$^{35}$P experiment done by S. Khan and
collaborators~\cite{Khan85}. The sum of the deduced spectroscopic
factors from the proton pickup reaction from the whole $sd$ states
amounts to $\sum C^2S \approx 7.9$. Within the 20\% uncertainties of
the method, this exhausts the complete spectroscopic shell model
strength of $\sum C^2S = 8$. Experimental occupancies of the $2
s_{1/2}$ and $1 d_{3/2}$ orbitals as reported in
Table~\ref{OCCUP_SM_SI} are 1.63 and 0.31, respectively. Note that the
small occupancy of the $1d_{3/2}$ state is due to correlations.

As already mentioned, $4p4h$ neutron excitations from the $\nu d_{3/2}$
to the $\nu f_{7/2}$ orbitals have been allowed in the present
SM calculations. This provides an estimate of the contribution of
neutron cross-shell excitations (across the $N=20$ shell closure).
Such contributions are found to be negligible ($< 2$\%). Indeed, both
$^{36}$S and $^{34}$Si ground states have quasi pure single-particle
wave functions. For $^{34}$Si [$^{36}$S], the $(\pi d_{5/2})^6
(\nu sd)^{12}$ [$(\pi d_{5/2})^6 (\pi s_{1/2})^2 (\nu sd)^{12}$]
configuration represents about 87\% [85\%] of the ground state
wave function.
\begin{table}[!t]
  \caption{\label{OCCUP_SM_SI} Same as Table~\ref{OCCUP_SM_O} for $^{36}$S
    and $^{34}$Si. The third column shows the experimental occupancies
    obtained in Ref.~\cite{Khan85} for $^{36}$S.}
  \begin{ruledtabular}
    \begin{tabular}{ccccc}
      Orbital        &  $^{36}$S   & $^{34}$Si   & $^{36}$S (Ref.~\cite{Khan85}) \\
      $\pi 1 d_{5/2}$ &   5.85     &   5.75      &     6.1(12)     \\
      $\pi 2 s_{1/2}$ &   1.88     &   0.09      &     1.63(32)    \\
      $\pi 1 d_{3/2}$ &   0.27     &   0.16      &     0.31(6)     \\
    \end{tabular}
  \end{ruledtabular}
\end{table}
\begin{figure}[!b]
\includegraphics[width=8.6cm]{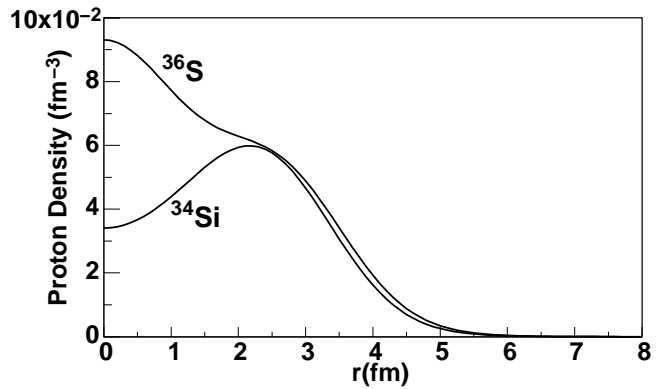}
\caption{\label{FIGURE2} Same as Fig.~\ref{FIGURE1} for proton densities
  in $^{36}$S and $^{34}$Si.}
\end{figure}
The mean occupation numbers for the proton orbitals deduced from the
present SM calculations are reported in Table~\ref{OCCUP_SM_SI}. The
agreement with the experimental values for $^{36}$S is very good,
lending confidence to the SM predictions for $^{34}$Si.  The study
of the $^{34}$Si(d,$^3He$)$^{33}$Al reaction channel should be used
in the future to determine if the occupancy of the $2 s_{1/2}$ proton
orbital has indeed dropped to nearly zero in $^{34}$Si.

The proton fraction removed from the $2 \pi s_{1/2}$
orbital, while moving from $^{36}$S to $^{34}$Si, is
significantly larger than the corresponding case in the
Oxygen isotopes.  Moreover, a larger mean occupation number
of the $d_{5/2}$ orbital\textbf{s} 
is predicted in $^{34}$Si as compared to
$^{22}$O. Both of these effects combined to explain the
larger depletion in the nuclear interior of $^{34}$Si, as
displayed in Fig.~\ref{FIGURE2}. The densities presented in
this figure have been obtained in the same way as for
$^{24,22}$O. The depletion factor in $^{34}$Si is found to
be F~=~43\%, significantly larger than in $^{22}$O. Thus,
we conclude that within this approach, $^{34}$Si is a good
candidate for a proton bubble. We are aware that the
nucleon densities reported above may not be as reliable as
the mean occupation numbers extracted from the SM code. The
self-consistent microscopic treatment presented in the
following sections is more appropriate to provide
radial densities.

\section{Mean Field calculations}

Self-consistent mean-field approaches enable to determine microscopically
the density distributions of nuclei. Nucleon occupation factors may also
be determined by taking pairing correlations into account. As a first step
in describing the density distributions, one must determine the role of
pairing correlations, if any, on the development of proton and neutron
bubbles in $^{34}$Si and $^{22}$O nuclei, respectively.

\subsection{Pairing effects}

As already alluded to in the Introduction, $^{22}$O is
expected to behave almost as a doubly-magic nucleus, being
that the $N=14$ sub-shell closure has been experimentally
determined. However, as
shown in the previous section, SM calculations predict a
17\% occupancy of the $2s$ neutron state, suggesting that
pairing correlations are likely to have some effect on this
nucleus. The effect of pairing correlations on the neutron
density profile of $^{22}$O will be shown in the following
subsections for both the non-relativistic and the
relativistic mean-field cases.

Let us now consider the case of $^{34}$Si. To start, we discuss the role
of pairing in the non-relativistic case. In this case pairing correlations
can be modeled in the Skyrme-Hartree-Fock-Bogoliubov (Skyrme-HFB) framework
by adopting the following zero-range density dependent pairing interaction:
\begin{equation}\label{eq:vpair}
 V_{pair}=V_0\left[1-\eta\left(\frac{\rho(r)}{\rho_0}\right)^\alpha\right]
 \delta\left({\bf r_1}-{\bf r_2}\right),
\end{equation}
with $\eta=0.5$ (mixed surface-volume interaction), $\alpha=1$ and
$\rho_0=0.16$ fm$^{-3}$. In the particle-hole channel, we employ the SLy4
Skyrme parametrization that is well suited to describe neutron-rich nuclei.

We fix the parameter $V_0$ in Eq.~(2) to reproduce the two-proton separation
energy in $^{34}$Si. Note that the two-proton separation energy is defined as:
 \begin{equation}
  S_{2p}=E(N,Z)-E(N,Z-2),
\end{equation}
where $E(N,Z)$ is the total binding energy of the ($N,Z$) nucleus.
It should be noted that the experimental value of 33.74 MeV
is already reasonably well reproduced without
pairing: the HF value is equal to 35.19 MeV. Moreover the
HFB calculations---which include the pairing
interaction---yield negligible corrections, as $Z=14$ is
predicted by the HFB approach to be a robust sub-shell
closure in agreement with the shell model spectroscopic
factors (see Table~II where the SM occupation of the $s$
state is only 4.5\%). Thus, we can safely perform the
analysis of this nucleus with the HF model, where pairing
is absent. Note that the above results will be confirmed in
Subsec.~C within the relativistic approach where pairing
effects are also found to collapse in the case of
$^{34}$Si.

\subsection{Non-relativistic mean field approach}

In this Subsection both $^{22}$O and $^{34}$Si are analyzed as possible
candidates for bubbles nuclei. Figure 3 displays neutron density
profiles in $^{22}$O (full line) and $^{24}$O (dashed line)
calculated self-consistently within the SLy4-HF approach.  In the case
of $^{24}$O the neutron single-particle state $2s_{1/2}$ is assumed to
be fully occupied.  The depletion of the central density in $^{22}$O
relative to $^{24}$O is clearly visible. However, the bubble profile
is not evident: since the central neutron density in $^{24}$O is
strongly enhanced, the depletion in $^{22}$O does not lead to the
development of a significant central hole.  The central depletion
fraction $F$ is $\sim$ 13\%, much weaker than the SM result. As one
switches on pairing and chooses the same parameters as in
Ref.~\cite{khan02} for the pairing interaction, the central hole is
seen to be completely washed out (see the dotted line in Fig.~3). Note
that the density profile of $^{24}$O remains unchanged when pairing is
switched on.

\begin{figure}
\begin{center}
\epsfig{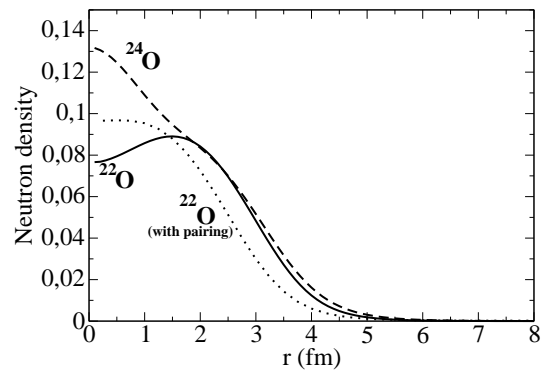}
\end{center}
\caption{HF neutron densities (in units of fm$^{-3}$) of $^{22}$O
(full line) and $^{24}$O (dashed line) calculated with the Skyrme
interaction SLy4. The dotted line represents the SLy4-HFB neutron density
of $^{22}$O.} \label{fig1}
\end{figure}

\begin{figure}[h]
\begin{center}
\epsfig{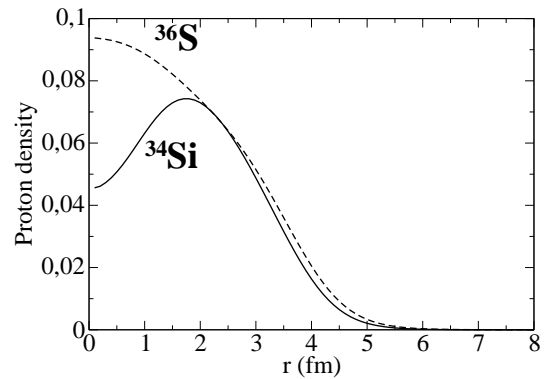}
\end{center}
\caption{HF proton densities (in units of fm$^{-3}$) of $^{36}$S
(dashed line) and $^{34}$Si (solid line) calculated with the
Skyrme interaction SLy4.} \label{fig2}
\end{figure}

The SLy4-HF proton density profiles calculated in $^{34}$Si
and $^{36}$S (where the $s$ state is fully occupied) are
shown in Fig. 4. One observes that the bubble is much more
prominent in this case than in $^{22}$O. The depletion
fraction $F$ is $\sim$~38\% and in very good agreement with
the SM value.  We should mention that pairing is expected
to modify the density profile of $^{36}$S, but this is not
relevant here since this density is plotted only to better
appreciate the central hole in $^{34}$Si. By comparing the
HF proton density in $^{34}$Si with the HF neutron density
in $^{22}$O, one observes that the central value in
$^{34}$Si is much lower than in $^{22}$O. The contribution
to the central value of the density is entirely due
to the first $s$ wave function, i.e. the 1$s$. 
The difference between the two
central values may be related to the presence of a neutron
excess at the surface of $^{34}$Si. The effect of this
neutron-skin on the proton $1s_{1/2}$ wave function is to
attract and push it towards the surface, thereby lowering
its value at the center. This can be observed in Fig. 5
where the neutron (proton) $1s$ contribution to the HF
density is plotted for $^{22}$O ($^{34}$Si).

\begin{figure}[h]
\begin{center}
\epsfig{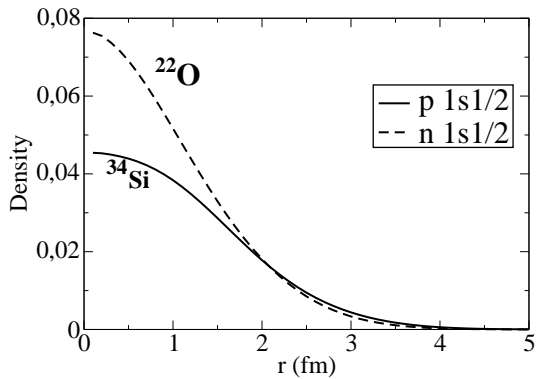}
\end{center}
\caption{Neutron (proton) $1s$ contributions to the density
(in units of fm$^{-3}$) for $^{22}$O ($^{34}$Si).} \label{figOSi}
\end{figure}

\subsection{Relativistic mean field approach}

As in the previous section, calculations are performed for the two
Oxygen isotopes $^{22}$O and $^{24}$O as well as for the two
$N\!=\!20$ isotones $^{34}$Si and $^{36}$S, but now using a
relativistic mean field (RMF) approach.  Pairing effects are
evaluated within the relativistic Hartree-Bogoliubov (RHB) model.
In one particular realization of 
the relativistic formalism the dynamics of the system is dictated
by an interacting Lagrangian density of the following form:
\begin{align}
{\mathscr L}_{\rm int} & =
 \bar\psi\left[g_{\rm s}\phi   \!-\!
         \left(g_{\rm v}V_\mu  \!+\!
    \frac{g_{\rho}}{2}\tau\cdot{\bf b}_{\mu}
                               \!+\!
    \frac{e}{2}(1\!+\!\tau_{3})A_{\mu}\right)\gamma^{\mu}
         \right]\psi \nonumber \\
                   & -
    \frac{\kappa}{3!} (g_{\rm s}\phi)^3 \!-\!
    \frac{\lambda}{4!}(g_{\rm s}\phi)^4 \!+\!
    \frac{\zeta}{4!}
    \Big(g_{\rm v}^2 V_{\mu}V^\mu\Big)^2 \nonumber \\
                   & \!+\!
    \Lambda_{\rm v}
    \Big(g_{\rho}^{2}\,{\bf b}_{\mu}\cdot{\bf b}^{\mu}\Big)
    \Big(g_{\rm v}^2V_{\mu}V^\mu\Big) \;.
 \label{Lagrangian}
\end{align}
where $\psi$ represents an isodoublet nucleon field interacting via
the exchange of two isoscalar mesons --- a scalar ($\phi$) and a
vector ($V^{\mu}$) --- one isovector meson ($b^{\mu}$), and the photon
($A^{\mu}$)~\cite{pi1,pi2}. In addition to meson-nucleon interactions,
the Lagrangian density is supplemented by non-linear meson
interactions with coupling constants denoted by $\kappa$, $\lambda$,
$\zeta$, and $\Lambda_{\rm v}$ that are responsible for a softening of
the nuclear-matter equation of state, both for symmetric and
pure-neutron matter.  For the RMF case we consider two
parametrizations: the very successful NL3 parameter set~\cite{pi3,pi4}
and a more recent set known as FSUGold~\cite{pi5}. The main difference
between these two models lies in the prediction of the density
dependence of the symmetry energy.  This difference manifests itself
in significantly larger neutron skins for NL3 than for FSUGold
\cite{pi5}. Neutron skins for the two isotones of interest in the
present work, alongside other ground-state properties, have been
listed in Table~\ref{Table1} for $^{34}$Si and $^{36}$S.

\begin{table}
\begin{tabular}{|c|c|c|c|}
 \hline
  Model & $B/A({\rm MeV})$ & $R_{ch}({\rm fm})$ &
  $R_{n}\!-\!R_{p}({\rm fm})$\\
 \hline
  NL3        & $8.36$ & $3.13$ & $0.25$ \\
  FSUGold    & $8.28$ & $3.13$ & $0.21$ \\
  Experiment & $8.34$ &  ---   &   ---  \\
 \hline
  NL3        & $8.50$ & $3.26$ & $0.12$ \\
  FSUGold    & $8.42$ & $3.26$ & $0.09$ \\
  Experiment & $8.58$ & $3.28$ &   ---  \\
\hline
\end{tabular}
\caption{Binding energy per nucleon, charge radii, and neutron
skin thickness for ${}^{34}$Si (upper block) and ${}^{36}$S (lower
block) as predicted by the two RMF models used in this work. When
available, experimental data is provided for comparison.}
\label{Table1}
\end{table}

RMF neutron densities for the two neutron-rich isotopes
$^{22}$O and $^{24}$O are displayed in Fig. 6. Whereas the
RMF results show a mild model dependence, differences
between the relativistic and non-relativistic models are
significant. Indeed, in contrast to the non-relativistic
case, the relativistic results display no enhancement of
the central neutron density in $^{24}$O. Moreover, the
removal of both $2s_{1/2}$ neutrons from $^{24}$O yields a
strong depletion of the interior neutron density in
$^{22}$O. As a result, a central depletion fraction of
$F\!=\!34\%(28\%)$ is predicted for $^{22}$O by the
FSUGold(NL3) model. These values are significantly larger
than the 13\% depletion fraction obtained with the SLy4-HF
parametrization, but close to the SM expectations.
It may be interesting to elucidate the source behind this
discrepancy which might be related to the saturation
mechanism in RMF and HF calculations.

\begin{figure}
\begin{center}
\vspace{0.2in}
\epsfig{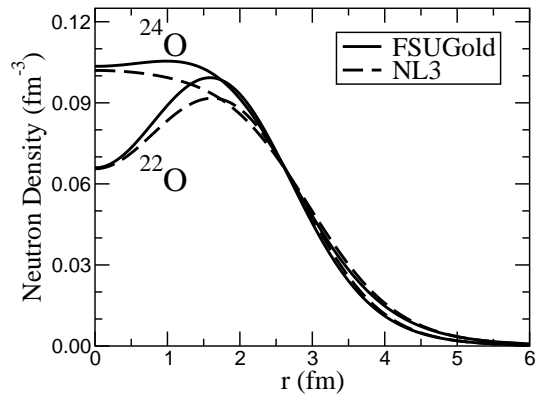}
\end{center}
\caption{RMF neutron densities of $^{22}$O and $^{24}$O
 calculated with the two RMF models described in the text.}
\label{fig3}
\end{figure}

In the case of the larger nuclei $^{34}$Si and $^{36}$S one
observes, now in agreement with the 
non-relativistic and the SM cases, how the proton density of
$^{34}$Si is significantly depleted in the nuclear interior
and how the proton bubble disappears as soon as the
$2s_{1/2}$ proton orbital is filled in $^{36}$S (see
Fig.~\ref{fig4}). Indeed, not only is the central density
in $^{34}$Si strongly depleted, but the central density in
$^{36}$S also appears to be get greatly enhanced. This
behavior results in central depletion factors for $^{34}$Si
of $F\!=\!42\%$ and $F\!=\!37\%$ for the FSUGold and NL3
parameter sets, respectively. These numbers are in good
agreement with the non-relativistic prediction of
$F\!\sim\!38\%$ and SM of $\sim$ 43\%.

\begin{figure}
\begin{center}
\vspace{0.2in}
\epsfig{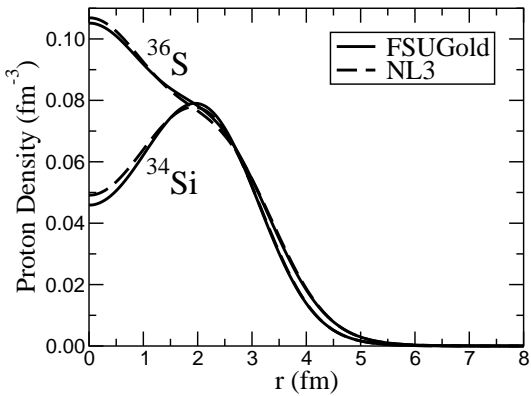}
\end{center}
\caption{RMF proton densities of $^{36}$S and $^{34}$Si
calculated with the two RMF models described in the text.}
\label{fig4}
\end{figure}

Let us quantify now the effects of pairing correlations within the RHB
model.  A medium dependence for a relativistic mean-field interaction
can either be introduced by including non-linear meson
self-interaction terms in the Lagrangian, as in the case of NL3 and
FSUGold, or by assuming an explicit density dependence for the
meson-nucleon couplings. This is the case of the DD-ME2
model~\cite{LNVR.05} that we adopt to perform RHB calculations. The
couplings of the $\sigma$-meson and $\omega$-meson to the nucleon are
assumed to be of the form:
\begin{equation}
 g_{i}(\rho) = g_{i}(\rho_{sat})f_{i}(x) \quad\text{for} \quad i=\sigma
,\omega\;, \label{gcoupl}%
\end{equation}
where
\begin{equation}
f_{i}(x) = a_{i}\frac{1+b_{i}(x+d_{i})^{2}}{1+c_{i}(x+d_{i})^{2}}
\label{fcoupl}%
\end{equation}
is a function of $x=\rho/\rho_{sat}$, and $\rho_{sat}$ denotes the nucleon
density at saturation in symmetric nuclear matter. Constraints at nuclear
matter saturation density and at zero density are used to reduce the number
of independent parameters in Eq.~(\ref{fcoupl}) to three. Three additional
parameters in the isoscalar channel are $g_{\sigma}(\rho_{sat}),\;
g_{\omega}(\rho_{sat})$, and $m_{\sigma}$---the mass of the phenomenological
$\sigma$ meson. For the $\rho$ meson coupling the functional form of the
density dependence is suggested by Dirac-Brueckner calculations of asymmetric
nuclear matter:
\begin{equation}
g_{\rho}(\rho)=g_{\rho}(\rho_{sat})\exp[-a_{\rho}(x-1)]\;,
\end{equation}
and the isovector channel is parametrized by $g_{\rho}(\rho_{sat})$ and
$a_{\rho}$. Bare values are used for the masses of the $\omega$ and $\rho$
mesons: $m_{\omega}=783$ MeV and $m_{\rho}=763$ MeV.
DD-ME2 is determined by eight independent parameters, 
adjusted to the properties of symmetric
and asymmetric nuclear matter, binding energies, charge radii, and neutron
radii of spherical nuclei~\cite{LNVR.05}. The interaction has been
tested in the calculation of ground state properties of large set of spherical
and deformed nuclei. When
used in the relativistic RPA, DD-ME2 reproduces with high accuracy data on
isoscalar and isovector collective excitations \cite{LNVR.05,PVKC.07}.

In Figs. \ref{dd-me2_O} and  \ref{dd-me2_Si} we display the neutron (proton)
density profiles for $^{22,24}$O, $^{34}$Si and $^{36}$S, calculated in RHB
model~\cite{VALR.05} with the DD-ME2 effective interaction in the particle-hole
channel, and with the Gogny interaction~\cite{BGG.84} in the pairing channel
\begin{align}
V^{pp}(1,2)~&=~\sum_{i=1,2}e^{-(({\bf r}_{1}-{\bf r}_{2})/{\mu _{i}}%
)^{2}}\,(W_{i}~+~B_{i}P^{\sigma } \nonumber \\ &-H_{i}P^{\tau }-M_{i}P^{\sigma }P^{\tau }),
\end{align}
with the set D1S \cite{BGG.91} for the parameters $\mu _{i}$, $W_{i}$,
$B_{i}$, $H_{i}$, and $M_{i}$ $(i=1,2)$.

\begin{figure}
\begin{center}
\vspace{0.2in}
\epsfig{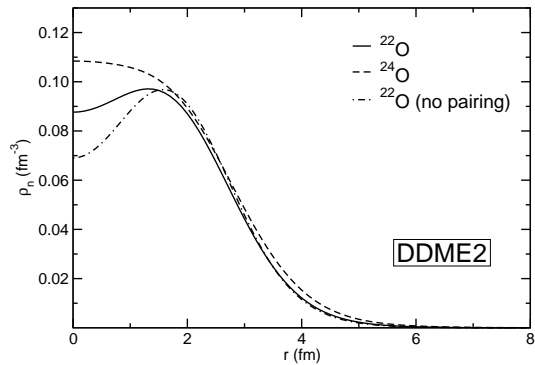}
\end{center}
\caption{Neutron density profiles of $^{22}$O and $^{24}$O
calculated in the RHB model with the density-dependent RMF interaction
DD-ME2, and Gogny pairing.
}
\label{dd-me2_O}
\end{figure}
\begin{figure}
\begin{center}
\vspace{0.2in}
\epsfig{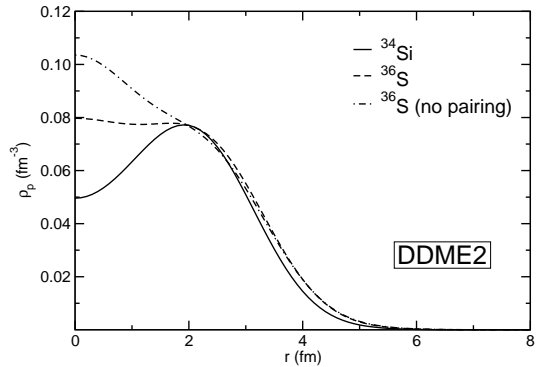}
\end{center}
\caption{Proton densities of $^{36}$S and $^{34}$Si
calculated in the
RHB model with the DD-ME2 interaction plus Gogny D1S pairing.}
\label{dd-me2_Si}
\end{figure}

For $^{24}$O and $^{34}$Si the RHB calculation with the DD-ME2
interaction predicts neutron and proton density profiles similar to those
calculated with NL3 and FSUGold.  Because of the large gaps: between
$\nu s_{1/2}$ and $\nu d_{3/2}$ ($> 4$ MeV) in $^{24}$O, and between
$\pi d_{5/2}$ and $\pi s_{1/2}$ ($> 6$ MeV) in $^{34}$Si, we find a pairing
collapse in these nuclei, in agreement with non-relativistic predictions.
On the other hand, the inclusion of pairing correlations has a pronounced
effect on the neutron and proton density profiles in $^{22}$O and $^{36}$S,
respectively. When pairing is set to zero (dash-dot curves in Figs.
\ref{dd-me2_O} and \ref{dd-me2_Si}) the $\nu s_{1/2}$ orbital is empty in
$^{22}$O, and the $\pi s_{1/2}$ orbital is fully occupied in $^{36}$S. The
resulting DD-ME2 density profiles are again very similar to those calculated
with the two other RMF interactions. However, the pairing interaction in the
RHB model calculation modifies the occupancy of the two $2s_{1/2}$ orbitals,
thus reducing the pronounced bubble in the neutron density of $^{22}$O, as
well as the prominent cusp in the proton density of $^{36}$S.

In the DD-ME2 model the F values are found equal to 29\%, 10\% and
36\% for $^{22}$O (without pairing), $^{22}$O (with pairing) and
 $^{34}$Si (same result with and without pairing), respectively.

\section{Summary and conclusions}

The occurrence of proton and neutron bubbles in $^{34}$Si and
$^{22}$O, respectively, has been investigated using three
different theoretical approaches: (i) the shell model,
(ii) the Skyrme mean-field model, and (iii) the relativistic
mean-field model. 
This occurrence can be quantified by the values of the depletion fraction 
$F$ which we have evaluated in these different approaches and which are
 summarized in table IV. 

\begin{table}
\begin{tabular}{|c|c|c|c|c|c|c|c|}
 \hline
  Nucleus & SM & SLy4  & SLy4 & NL3  & FSUGold  &
DDME2  & DDME2 \\
          &    & HF & HFB & RMF & RMF &  RMF & RHB \\
\hline
  $^{22}$O   & 25\% & 13\% & 0 & 28\% & 34\% & 29\% & 10\% \\
\hline
  $^{34}$Si  & 43\% & 38\% & 38\% & 37\% & 42\% & 36\% & 36\% \\
\hline
\end{tabular}
\caption{Central fraction of depletion $F$}
\label{Table4}
\end{table}

For the $^{22}$O nucleus a significant model dependence has been found.
Moreover, in both non-relativistic and relativistic cases, pairing
correlations have been shown to weaken the bubble phenomenon. In
contrast, for $^{34}$Si an overall agreement exists: a central
depletion fraction of $\sim$ 40\% is predicted by all the models.
Although not discussed here, the presence of this proton bubble is
expected to induce quenching of the spin-orbit splitting of the
neutron $2p$ orbitals in $^{34}$Si (see Refs.~\cite{tod04,Gaud07}
for details).

These robust results indicate that $^{34}$Si is a good candidate for
a bubble density profile. The measurement of the charge density in
$^{34}$Si could be undertaken, for instance, by electron scattering
in a exotic beam collider, such as EXL in FAIR and RIBF in Riken.
The bubble impact on the momentum distribution in these experiment
has been investigated in Ref. \cite{kha08}. The study of $^{34}$Si,
either by high energy proton scattering (to focus on the matter
distribution) or by direct reactions (to yield the spectroscopic
factors and the low-energy excitation spectrum) is already
feasible~\cite{kha08}. For the transfer reaction, the
$^{34}$Si($d$, $^{3}$He)$^{33}$Al reaction channel should be used to
determine if the occupancy of the $2s_{1/2}$ proton orbit has dropped
to nearly zero in $^{34}$Si. This would confirm the SM predictions
shown in Sec.~II, while providing strong evidence in favor of a
strongly depleted central density in $^{34}$Si.

\vskip 0.5cm

\noindent{\bf Acknowledgments}
The authors thank K. Yoshida for valuable discussions.
The research of J.P. is supported in part by the United States
DOE grant DE-FD05-92ER40750.

\newpage

\end{document}